\documentstyle[aps,amsfonts,preprint,tighten,epsfig]{revtex}

\begin{document}

\newcommand{\order}{O}
\newcommand{\Lie}{\pounds}
\newcommand{\reals}{{\Bbb R}}
\newcommand{\ball}{{\cal B}}
\newcommand{\sphere}{{\cal S}}
\newcommand{\fEM}{f_{\rm EM}}
\newcommand{\fS}{f_{\rm S}}
\newcommand{\fStil}{\tilde{f}_{\rm S}}
\newcommand{\gS}{g_{\rm S}}
\newcommand{\gStil}{\tilde{g}_{\rm S}}
\newcommand{\fext}{f_{\rm ext}}
\newcommand{\TS}{T_{\rm S}}
\newcommand{\Tbody}{T_{\rm body}}
\newcommand{\Text}{T_{\rm ext}}
\newcommand{\TEM}{T_{\rm EM}}
\newcommand{\phione}{\phi}
\newcommand{\phitwo}{\bar{\phi}}
\newcommand{\phipm}{\phi_\pm}
\newcommand{\phip}{\phi_+}
\newcommand{\phim}{\phi_-}
\newcommand{\nablatil}{\widetilde{\nabla}}
\newcommand{\phitil}{\tilde{\phi}}
\newcommand{\phipmtil}{\tilde{\phi}_\pm}
\newcommand{\phiptil}{\tilde{\phi}_+}
\newcommand{\phimtil}{\tilde{\phi}_-}
\newcommand{\phiin}{\phi_{\rm in}}
\newcommand{\phitail}{\phi_{\rm tail}}
\newcommand{\step}{\theta}
\newcommand{\steppm}{\step_\pm}
\newcommand{\stepp}{\step_+}
\newcommand{\stepm}{\step_-}
\newcommand{\taupm}{\tau_\pm}
\newcommand{\taup}{\tau_+}
\newcommand{\taum}{\tau_-}
\newcommand{\taubarpm}{T_\pm}
\newcommand{\taubarp}{T_+}
\newcommand{\taubarm}{T_-}
\newcommand{\tausig}{\tau_\Sigma}
\newcommand{\Gone}{G^{(1)}}
\newcommand{\Gpm}{G_\pm}
\newcommand{\Gp}{G_+}
\newcommand{\Gm}{G_-}
\newcommand{\sigdot}{\dot{\sigma}}
\newcommand{\sigddot}{\ddot{\sigma}}
\newcommand{\Udot}{\dot{U}}
\newcommand{\Uddot}{\ddot{U}}
\newcommand{\Adot}{\dot{A}}
\newcommand{\rhat}{\hat{r}}
\newcommand{\gbar}{\bar{g}}
\newcommand{\adot}{\dot{a}}
\newcommand{\Ptil}{\widetilde{P}}
\newcommand{\util}{\tilde{u}}
\newcommand{\atil}{\tilde{a}}
\newcommand{\al}{\alpha}
\newcommand{\be}{\beta}
\newcommand{\ga}{\gamma}
\newcommand{\de}{\delta}
\newcommand{\ep}{\epsilon}

\title{Axiomatic approach to radiation reaction of scalar point
particles in curved spacetime}

\author{Theodore C. Quinn}
\address{Enrico Fermi Institute and Department of Physics\\
The University of Chicago\\
5640 S. Ellis Avenue\\
Chicago, Illinois 60637-1433}

\date{\today}

\maketitle

\begin{abstract}

Several different methods have recently been proposed for calculating
the motion of a point particle coupled to a linearized gravitational
field on a curved background. These proposals are motivated by the
hope that the point particle system will accurately model certain
astrophysical systems which are promising candidates for observation
by the new generation of gravitational wave detectors. Because of its
mathematical simplicity, the analogous system consisting of a point
particle coupled to a scalar field provides a useful context in which
to investigate these proposed methods. In this paper, we generalize
the axiomatic approach of Quinn and Wald in order to produce a general
expression for the self force on a point particle coupled to a scalar
field following an arbitrary trajectory on a curved background. Our
equation includes the leading order effects of the particle's own
fields, commonly referred to as ``self force'' or ``radiation
reaction'' effects. We then explore the equations of motion which
follow from this expression in the absence of non-scalar forces.

\end{abstract}

\section{Introduction}
\label{intro}

There has been much recent interest in calculating the motion of
astrophysical systems which emit gravitational waves in anticipation
of data from a new generation of detectors. Full three-dimensional
numerical simulations are required in order to produce useful results
for many of the most promising observational candidates, such as
colliding black holes. However, there also exists a large class of
systems which can be accurately modelled by a small isolated body
moving in the fixed background created by a much larger body (e.g.,\ a
solar mass star falling into a supermassive black hole). For such a
system, we might hope to produce useful results by treating the
smaller object as a point particle and introducing the effects of its
fields and internal structure as perturbations to the background
geodesic orbit.

The perturbations due to the particle's own fields, commonly called
``radiation reaction'' or ``self force'' effects, are particularly
important because they include the forces responsible for the decay of
the body's orbit. If both the background spacetime and the unperturbed
orbit of the body possess enough symmetry, it is possible to infer the
effects of these forces on the orbit from global conservation
principles: one calculates the energy and/or angular momentum radiated
to infinity by a particle in geodesic motion, and then modifies the
orbit to reflect this energy and angular momentum loss in a
time-averaged fashion. (Obviously, this procedure can be iterated if
greater accuracy is required.) Some justification for this method is
provided by Quinn and Wald~\cite{quinnwaldconservation}. However, in
the absence of such symmetries, it is necessary to directly calculate
the effects of the local fields in the neighborhood of the
particle. Unfortunately, this problem is ill-posed, since the fields
diverge in the neighborhood of the particle's world line, so that any
such local calculation must include a rule for extracting the
appropriate finite part of these divergent fields.

There is an extensive literature devoted to this regularization
problem. In 1938, Dirac~\cite{dirac} reproduced the force expression
(originally given by Abraham~\cite{abraham}) for a point particle
coupled to an electromagnetic field in Minkowski spacetime by imposing
local energy conservation on a tube surrounding the particle's world
line and subtracting the infinite contributions to the force through a
``mass renormalization'' scheme.  In 1960, Dewitt and
Brehme~\cite{dewittbrehme} generalized this approach to an arbitrary
curved background spacetime. (A trivial calculational error in their
paper was later corrected by Hobbs~\cite{hobbs}.) More recently, Mino
et al.~\cite{mino} further adapted this approach to produce a force
expression for a point particle coupled to a linearized gravitational
field on a vacuum background spacetime, and Quinn and
Wald~\cite{quinnwaldforce} rederived both the electromagnetic and
gravitational forces using an axiomatic approach which, in effect,
regularizes the forces by comparing forces in different spacetimes.

There has emerged from this work a consensus regarding the correct
equation of motion for a particle coupled to electromagnetic fields on
an arbitrary curved background and for a particle coupled to
linearized gravitational fields on a vacuum background. In principle,
the latter equation allows one to calculate the dynamics of the
astrophysical systems of interest described above. In practice,
however, very little progress has been made in applying either
equation of motion to concrete physical examples for two
reasons. First, given a world line in an arbitrary spacetime, the
calculation of the associated retarded fields is a complex and
difficult problem. Second, once these fields are calculated, the
equations of motion require one to identify that portion of the
retarded field at each point of the world line which arises from
source contributions interior to the light cone. This part of the
field is often called the ``tail term,'' and most approximation
schemes for calculating the retarded field entangle the tail and
non-tail contributions to the field.

Nevertheless, some progress has been made, notably in the
electromagnetic case. In 1964, DeWitt and DeWitt~\cite{dewittdewitt}
calculated the tail term for an electromagnetic particle in a circular
orbit on a Schwarzschild background to leading order in the background
curvature and the velocity of the particle. In 1980, Smith and
Will~\cite{smithwill} calculated the force on an electromagnetic
particle held static on a Schwarzschild background, essentially by
repeating DeWitt and Brehme's local stress-energy conservation
argument. Neither result has been generalized to the case of a massive
particle coupled to gravitational fields, nor has there been any
direct progress on the more complex systems which are of interest to
the gravitational wave astronomy community. However, several new ideas
have emerged in recent years which may lead to further
progress. Ori~\cite{ori} has suggested an alternative regularization
scheme involving averaging of multipole moments which is better
adapted to concrete calculations, while others have suggested a hybrid
scheme in which the tail term is calculated through a combination of
Hadamard expansion techniques for small distances and multipole
techniques for larger distances~\cite{wisemanprivate}.

It is clearly important to test these ideas. In particular, we must
know whether these schemes are equivalent to the equations of motion
discussed above. Because of its mathematical simplicity, one natural
system in which to explore all of these questions is that of a point
particle coupled to a scalar field. Motivated by this, several
researchers have begun to apply the ideas discussed above to the
scalar system. In particular, Ori's method has been applied to the
motion of scalar particles in the Kerr spacetime~\cite{orikerr} and in
the Schwarzschild spacetime~\cite{burko,barackori}, and
Wiseman~\cite{wisemanstatic} has adapted the calculation of Smith and
Will in order to calculate the force on a scalar particle held static
in the Schwarzschild spacetime. In the present paper, we generalize
the axiomatic approach of Quinn and Wald~\cite{quinnwaldforce} in
order to produce the general equation of motion for a point particle
coupled to a scalar field on an arbitrary background spacetime. It is
hoped that this general expression will be useful in evaluating the
validity of the calculational schemes described above for the scalar
case, and that this comparison will ultimately help to clarify the
relationship between the various methods which have been proposed for
the electromagnetic and gravitational cases.

In Sec.~\ref{force}, we derive an expression for the force on a
particle following an arbitrary trajectory in curved spacetime. Then,
in Sec.~\ref{eom}, we explore the equations of motion which follow
from this expression in the absence of non-scalar forces.

\section{The scalar force}
\label{force}

Given a spacetime containing a particle world line and a Klein-Gordon
field sourced by the particle, we wish to define the total scalar
force $\fS^a$ on the particle at each point of the world line,
including so-called self-force or radiation reaction effects. For an
electromagnetic point particle in flat spacetime, an expression of
this sort was first given by Abraham~\cite{abraham} in 1905, was later
rederived in a relativistic context by Dirac~\cite{dirac}, and is
often found in textbooks (e.g.,\ Jackson~\cite{jackson}). However,
since there are no classical point particles in nature, and the
theoretical status of such objects is problematic at best, it is
important to ask how any such prescription is constrained by physics.

Our view is that the force law should reflect the force on an extended
body coupled to a scalar field in the limit of small spatial
extent. In particular, fix the background spacetime and consider a
family of extended bodies and corresponding scalar fields
parameterized by $\epsilon$, the spatial size of the bodies. For each
body in the family, we define a center of mass world line $z(\tau)$
(e.g., by the methods of Beiglb\"{o}ck~\cite{beiglbock}) and calculate
the charge, $q$, and mass, $m$, of the body with respect to this world
line,\footnote{Because the scalar charge density is a scalar quantity,
the total charge that one calculates for an extended body depends upon
the spacelike surface used to slice the body. This is in contrast to
electromagnetism, where the charge density is the time component of a
conserved vector field and the total charge is independent of slice.}
as well as the force $\fS^a[\epsilon]$ exerted by the scalar field on
the body. (For the definition of the force exerted on a small body by
a field to which it is coupled, see Quinn and
Wald~\cite{quinnwaldforce}.) We further require that $m$ and $q$
vanish as $\epsilon$ goes to zero. For such a one-parameter family,
Quinn and Wald~\cite{quinnwaldforce} argue that it is possible to
specify some set of conditions on the internal structure and
composition of the extended bodies such that the limit of
$\fS^a[\epsilon]$ for small $\epsilon$ is independent of their
internal details. We would like our expression for $\fS^a$ to
correctly the describe the order $q$ and $q^2$ contributions to
$\fS^a[\epsilon]$ which are independent of the internal structure of
the body under these conditions. (Other corrections which arise from
the internal structure, such as multipole effects and spin effects,
have been derived elsewhere and should simply contribute additively at
this order.)

Unfortunately, the limit described above is quite delicate, and the
task of specifying conditions to ensure its convergence appears to be
formidable. (The analysis of Dixon~\cite{dixon} demonstrates the
degree of complexity which arises even without considering self-field
effects.) Nevertheless, certain properties of this limit are strongly
suggested by the nature of the divergences in the scalar
field. Following Quinn and Wald~\cite{quinnwaldforce}, we will
introduce these properties as axioms, and then give the unique
prescription for $\fS^a$ which satisfies these axioms.

In the next subsection, we will motivate our crucial Comparison Axiom
by considering the point particle limit described above and develop
the expansions required to state the axiom. Then, in the following
subsection, we state both axioms and give the unique prescription for
$\fS^a$ that satisfies them, which is the main result of this paper.

\subsection{Motivation for the Comparison Axiom}
\label{motivation}

Consider a spacetime $(M, g_{ab})$ containing a spatially compact body
characterized by stress-energy $\Tbody^{ab}$ and scalar charge density
$\rho$, a smooth Klein-Gordon field $\phi$, and
possibly some other set of fields which are coupled to the body,
characterized by $\Text^{ab}$. The Klein-Gordon field $\phi$ satisfies
the equation 
\begin{equation}
\nabla^a \nabla_a \phi = - 4 \pi \rho
\label{kg}
\end{equation}
with stress-energy
\begin{equation}
\TS^{ab} = \frac{1}{4\pi}(\nabla^a \phi \nabla^b \phi
                 - \frac{1}{2} g^{ab} g_{cd} \nabla^c \phi \nabla^d \phi).
\end{equation}
Assuming that the total stress-energy is conserved, so that
\begin{equation}
\nabla_b(\Tbody^{ab} + \TS^{ab} + \Text^{ab}) = 0,
\end{equation}
then the force density exerted on the body by the scalar field is
given by
\begin{equation}
\nabla_b \Tbody^{ab} + \nabla_b \Text^{ab}
  = - \nabla_b \TS^{ab}
  = \rho \nabla^a \phi.
\label{forcedensity}
\end{equation}

Therefore, naively taking the point particle limit, we would expect
the force on a scalar particle of charge $q$ to be given by
\begin{equation}
\fS^a = q \nabla^a \phi.
\label{fSnaive}
\end{equation}
Unfortunately, this expression is meaningless as it stands, since
$\nabla_a \phi$ diverges on the world line of the particle. (The
situation is exactly the same with the Lorentz force law $\fEM^a = q
F^{ab} u_b$.) However, if we consider two points $P$ and $\Ptil$ along
the world lines of two different particles in two different spacetimes
(each with charge $q$), and we identify the neighborhoods of $P$ and
$\Ptil$, then we might hope that, under some conditions, the
difference $\nabla^a \phi - \nablatil^a \phitil$ will be finite even as
the two individual fields diverge. Under such conditions, it seems
reasonable to expect that the difference between the forces on the
particles will be given by the (finite) difference between the field
gradients. That is,
\begin{equation}
\fS^a - \fStil^a = \lim_{r \rightarrow 0} q
                   \langle \nabla^a \phi - \nablatil^a \phitil \rangle_r.
\label{subtract}
\end{equation}
(Here, the average over a sphere of radius $r$, denoted by $\langle
\rangle_r$, is introduced to allow for the possibility that the $r
\rightarrow 0$ limit of the difference is finite, but
direction-dependent.)

Quinn and Wald~\cite{quinnwaldforce} give plausibility arguments which
suggest that the counterpart of Eq.~(\ref{subtract}) is indeed a
property of the point particle limit in the electromagnetic
and gravitational cases. These arguments generalize straightforwardly
to the scalar case, so we will not give the details here. Instead, we
will simply impose Eq.~(\ref{subtract}) as an axiom and investigate
the consequences for $\fS^a$. This idea will be the basis of our
crucial Comparison Axiom in the next subsection. However, first we
must find out what conditions to impose on the spacetimes, the world
lines near $P$ and $\Ptil$, and the identification of their
neighborhoods in order to ensure that the difference in the field
gradients be finite as $r \rightarrow 0$. In order to answer this
question, we will now examine in detail the singularity structure of
the scalar field in the neighborhood of the world line.

Consider a scalar field satisfying Eq.~(\ref{kg}) in a spacetime $(M,
g_{ab})$ with a point particle source
\begin{equation}
\rho(x) = \int \!\! q \delta^4(x,z(\tau)) \, d\tau.
\end{equation}
In contrast to the electromagnetic case, the Klein-Gordon equation
does not require conservation of charge. For simplicity, we shall
assume throughout our analysis that the charge $q$ is constant along
the world line. We wish to expand $\phi$ in $r$, the spatial distance
from the world line $z(\tau)$. We are primarily interested in the
divergent contributions to $\phi$, characterized by the negative
powers of $r$ in the expansion, since these divergent contributions
will determine the conditions required for convergence of the limit in
Eq.~(\ref{subtract}). It follows from the general theory of
propagation of singularities (see theorem 26.1.1 of
Hormander~\cite{hormander}) that every solution of Eq.~(\ref{kg})
which is smooth away from the world line will have the same
singularity structure near the world line, so we are free to choose
any convenient solution for our expansion. Later, when we wish to
produce an explicit expression for $\fS^a$, we will want to write
$\phi$ in terms of the advanced and retarded solutions. Therefore,
these are the solutions which we will analyze in the following
expansion.

Given any point $x$ in a spacetime $(M,g_{ab})$, there exists a convex
normal neighborhood $C(x)$ containing $x$ [i.e.\ a neighborhood $C(x)$
such that there exists a unique geodesic connecting any two points
within $C(x)$]. For $x' \in C(x)$, the Hadamard elementary solution of
Eq.~(\ref{kg}) can be written in the form~\cite{dewittbrehme}
\begin{equation}
\Gone(x,x') = \frac{1}{\pi}\left[
                \frac{U(x,x')}{\sigma(x,x')} + V(x,x')\ln|\sigma(x,x')|
                + W(x,x')\right],
\end{equation}
with corresponding advanced (+) and retarded (-) Green's functions
\begin{equation}
\Gpm(x,x') = \steppm(x,x')
             \Bigl[U(x,x') \delta\Bigl(\sigma(x,x')\Bigr)
                    - V(x,x') \step\Bigl(-\sigma(x,x')\Bigr)\Bigr].
\end{equation}
Here, $\sigma(x,x')$ is the biscalar of squared geodesic
distance\footnote{The biscalar of squared geodesic distance
$\sigma(x,x')$ is equal to half of the squared length of the geodesic
connecting $x$ and $x'$: negative for timelike separated events,
positive for spacelike separated events, and zero for null separated
events. It is defined only when there is a unique geodesic connecting
$x$ and $x'$.}  and $U$, $V$, and $W$ are all smooth biscalar
fields. (For an explanation of the bitensor formalism, see Dewitt and
Brehme~\cite{dewittbrehme}.)  The scalar function $\steppm(x,x')$ is
unity when $x'$ is in the causal future/past of $x$ and vanishes
otherwise.

\begin{figure}
\begin{center}
\epsfig{file=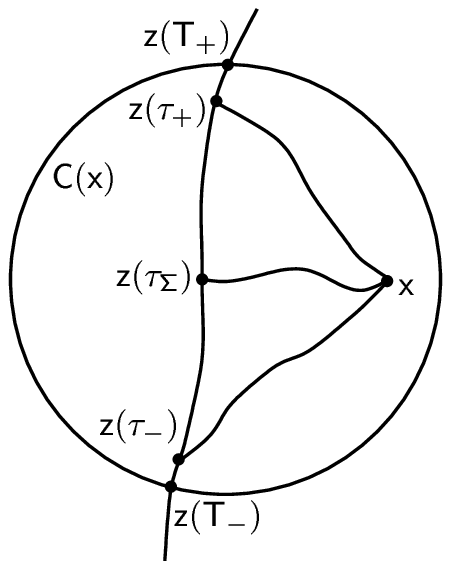}
\end{center}
\caption{\label{smallr} The neighborhood containing $x$ and $x'$.}
\end{figure}

For $x$ near the world line $z(\tau)$, let $\tausig$ be the proper
time along the world line which is simultaneous with $x$ in the sense
that the spatial surface $\Sigma$ generated by geodesics perpendicular
to $u^a$ at $z(\tausig)$ intersects $x$. In particular, let $x$ lie a
proper distance $r$ along the geodesic generated by unit spatial
vector $\rhat^a$ at $z(\tausig)$, and let $z(\taup)$ and $z(\taum)$ be
the intersection of the world line with the future and past light
cones of $x$, respectively. We require that $x$ be close enough to the
world line that $z(\tausig)$, $z(\taup)$, and $z(\taum)$ all lie
within the neighborhood $C(x)$, and we denote the future and past
intersections of the world line with the boundary of $C(x)$ by
$z(\taubarp)$ and $z(\taubarm)$, respectively. This is illustrated in
Fig.~\ref{smallr}. For the retarded field $\phim$, we then have
\begin{eqnarray}
\phim(x) &=& \int \!\! \Gm(x,x') \rho(x') \sqrt{-g} \, d^4x'\nonumber\\
  &=& \int \!\! \Gm(x,x')
                \left(\int \!\! q \delta^4(x', z(\tau)) \,d\tau \right)
                \sqrt{-g} \, d^4x' \nonumber\\
  &=& q \int \!\! \Gm(x,z(\tau)) \, d\tau \nonumber\\
  &=& q \int_{\taubarm}^{\taubarp} \!\!
        \stepm[x,z(\tau)]
        \Bigl[U(x,z(\tau)) \delta\Bigl(\sigma(x,z(\tau))\Bigr)
              - V(x,z(\tau)) \step\Bigl(-\sigma(x,z(\tau))\Bigr)\Bigr]
        \, d\tau \nonumber \\
   && \mbox{} + q \int_{-\infty}^{\taubarm} \!\!
                    \Gm(x,z(\tau)) \, d\tau \nonumber\\
  &=& q
      \int_{\taubarm}^{\tausig} \!\!
                \left[U \delta(\sigma)
                      - V \step(-\sigma)\right]
                \, d\tau
      + q \int_{-\infty}^{\taubarm}\!\!
              \Gm \, d\tau
\end{eqnarray}
In the last line and hereafter, we suppress the spacetime dependence
for all biscalars, since each depends upon $x$ in its first argument
and $z(\tau)$ in its second argument. For a bitensor $A$, we introduce
the notation
\begin{equation}
\dot{A} \equiv \frac{d}{d\tau}A(x,z(\tau))
        = u^{a'}\nabla_{a'}A(x,z(\tau)).
\end{equation}
We have
\begin{equation}
d\tau = \frac{d\tau}{d\sigma}d\sigma
      = \left(\frac{d\sigma}{d\tau}\right)^{-1} d\sigma
      = \sigdot ^{-1} d\sigma,
\end{equation}
which gives us
\begin{equation}
\phim
  =  q \left[
    \left\{\sigdot^{-1} U \right\}_{\tau=\taum}
    - \int_{\taubarm}^{\taum} \!\! V \, d\tau
    \right] 
    + q \int_{-\infty}^{\taubarm} \!\!
            \Gm \, d\tau.
\label{phimuv}
\end{equation}

We now wish to produce the corresponding expression for
$\nabla_a\phim$. Note that the right side of Eq.~(\ref{phimuv})
depends upon $x$ in two ways: explicitly through the first argument
of each biscalar and implicitly through $\taum$. 
We have
\begin{eqnarray}
\nabla_a \phim
  &=&  q \nabla_a \left[
      \left\{\sigdot^{-1} U \right\}_{\tau=\taum}
      - \int_{\taubarm}^{\taum} \!\! V \, d\tau
      \right]
      + q \int_{-\infty}^{\taubarm} \!\!
              \nabla_a \Gm \, d\tau \nonumber\\
  &=& q \biggl[
      \left\{-\sigdot^{-2} \nabla_a \sigdot U
               + \sigdot^{-1} \nabla_a U \right\}_{\tau=\taum}
      + \left\{-\sigdot^{-2} \sigddot U
               + \sigdot^{-1}\Udot\right\}_{\tau=\taum} \nabla_a\taum
\nonumber\\ && \phantom{\frac{q}{4\pi} \biggl[} \mbox{}
      - \int_{\taubarm}^{\taum} \!\! \nabla_a V \, d\tau
      - \{V\}_{\tau=\taum} \nabla_a\taum
      \biggr] 
      - q \int_{-\infty}^{\taubarm} \!\!
              \nabla_a \Gm \, d\tau
\label{gradphimraw}
\end{eqnarray}
Since $\sigma(x,z(\taum))=0$, we have
\begin{equation}
\nabla_a\{\sigma\}_{\tau=\taum}
  = \{\nabla_a\sigma\}_{\tau=\taum}
    + \{\sigdot\}_{\tau=\taum}\nabla_a\taum
  = 0,
\end{equation}
so that
\begin{equation}
\nabla_a\taum = \{-\sigdot^{-1} \nabla_a\sigma\}_{\tau=\taum}.
\end{equation}
Therefore, we have
\begin{eqnarray}
\nabla_a \phim
  &=& q \biggl[
     \left\{- \sigdot^{-2} \nabla_a \sigdot U
            +\sigdot^{-1} \nabla_a U 
            + \sigdot^{-3} \sigddot U \nabla_a \sigma
            - \sigdot^{-2} \Udot \nabla_a \sigma
            + \sigdot^{-1} V \nabla_a \sigma \right\}_{\tau=\taum}
\nonumber\\ && \phantom{\frac{q}{4\pi} \biggl[} \mbox{}
      - \int_{\taubarm}^{\taum} \!\! \nabla_a V \, d\tau
      \biggr] 
      + q \int_{-\infty}^{\taubarm} \!\!
              \nabla_a \Gm \, d\tau \nonumber\\
\label{gradphimuv}
\end{eqnarray}

In Eqs.~(\ref{phimuv}) and (\ref{gradphimuv}), we would like to
combine the integrals which appear on the right side. For
$\taubarm \leq \tau < \taum$, we have $G(x,z(\tau)) = -
V(x,z(\tau))$. Furthermore, since $V$ is a smooth biscalar,
\begin{equation}
\int_{\taubarm}^{\taum} \!\! V \, d\tau
  = \lim_{\ep \rightarrow 0} \int_{\taubarm}^{\taum - \ep} \!\! V \, d\tau
\end{equation}
and
\begin{equation}
\int_{\taubarm}^{\taum} \!\! \nabla_a V \, d\tau
  = \lim_{\ep \rightarrow 0} \int_{\taubarm}^{\taum - \ep} \!\!
                            \nabla_a V \, d\tau.
\end{equation}
Therefore, combining the integrals, we have
\begin{equation}
\phim
  = q
    \left\{\sigdot^{-1} U \right\}_{\tau=\taum}
    + \lim_{\epsilon \rightarrow 0}
         q \int_{-\infty}^{\taum - \epsilon} \!\!
            \Gm \, d\tau.
\label{phimuvtail}
\end{equation}
and
\begin{eqnarray}
\nabla_a \phim
  &=& q 
     \left\{- \sigdot^{-2} \nabla_a \sigdot U
            +\sigdot^{-1} \nabla_a U 
            + \sigdot^{-3} \sigddot U \nabla_a \sigma
            - \sigdot^{-2} \Udot \nabla_a \sigma
            + \sigdot^{-1} V \nabla_a \sigma \right\}_{\tau=\taum}
\nonumber\\ && \mbox{}
     +\lim_{\epsilon \rightarrow 0}
         q \int_{-\infty}^{\taum - \epsilon} \!\!
             \nabla_a \Gm \, d\tau \nonumber\\
\label{gradphimuvtail}
\end{eqnarray}

In order to investigate the singularity structure of $\phim$ and
$\nabla_a\phim$ near the world line, we need expansions for the
expressions in brackets on the right sides of
Eqs.~(\ref{phimuvtail}) and~(\ref{gradphimuvtail}) which are valid to
$\order[r^0]$. (The integrals in these equations make smooth
contributions to the fields.) The required small distance expansions
for $U$, $V$, $\sigma$, and their derivatives can all be found in
DeWitt and Brehme~\cite{dewittbrehme} or derived straightforwardly
from expressions given therein. Switching the roles of the primed and
unprimed indices for notational simplicity and including the
corresponding results for the advanced field, $\phip$, we have
\begin{equation}
\phipm(x') = q \left(r^{-1} - \frac{1}{2}a^{a}\rhat_{a}\right)
         \pm \lim_{\epsilon \rightarrow 0}
             q \int_{\taupm \pm \epsilon}^{\pm\infty} \!\!
                 \Gpm(x',z(\tau)) \, d\tau
         + \order[r]
\label{phipm}
\end{equation}
and
\begin{eqnarray}
\nabla_{a'}\phipm(x')
  &=& q \gbar_{a'a} \left(
      - r^{-2} \rhat^{a}
      - \frac{1}{2} r^{-1} a^{a}
      + \frac{1}{2} r^{-1} (a^{b} \rhat_{b}) \rhat^{a}
      - \frac{3}{8} (a^{b} \rhat_{b})^2 \rhat^{a}
      + \frac{3}{4} (a^{b} \rhat_{b}) a^{a}
\right. \nonumber \\ && \mbox{}
      - \frac{1}{6} R_{bdce} u^{b} u^{c} \rhat^{d} \rhat^{e} \rhat^{a}
      - \frac{1}{8} a^2 \rhat^{a}
      - \frac{1}{12} R_{bc} \rhat^{b} \rhat^{c} \rhat^{a}
      + \frac{1}{2} (\adot^{b} \rhat_{b}) u^{a}
\nonumber \\ && \mbox{}
      + \frac{1}{12} R_{bc} u^{b} u^{c} \rhat^{a}
      + \frac{1}{6} R_{bc} u^{b} \rhat^{c} u^{a}
      + \frac{1}{3} R^{a}{}_{cbd} u^{b} u^{c} \rhat^{d}
      \pm \frac{1}{3} a^2 u^{a}
      \mp \frac{1}{3} \adot^{a}
\nonumber \\ && \left. \mbox{}
      \mp \frac{1}{6} R_{bc} u^{b} u^{c} u^{a}
      + \frac{1}{6} R^{ab} \rhat_{b}
      - \frac{1}{12} R \rhat^{a}
      \mp \frac{1}{6} R^{ab} u_{b}
      \pm \frac{1}{12} R u^{a}
\right) \nonumber \\ && \mbox{}
     \pm \lim_{\epsilon \rightarrow 0}
         q \int_{\taupm\pm\epsilon}^{\pm\infty} \!\!
             \nabla_a \Gpm(x',z(\tau)) \, d\tau
     + \order[r],
\label{gradphipm}
\end{eqnarray}
where $\gbar_{a'a}$ is the bivector of geodetic parallel displacement,
defined by DeWitt and Brehme~\cite{dewittbrehme}.

We began this calculation in order to investigate what conditions we
need to impose on the spacetime neighborhoods and trajectories of
scalar particles in different spacetimes and on our identification of
these neighborhoods in order to ensure that the subtraction of field
gradients in Eq.~(\ref{subtract}) is finite, and Eq.~(\ref{gradphipm})
provides the answer to this question. Since the divergent terms in
Eq.~(\ref{gradphipm}) depend only upon the four-velocity and
four-acceleration of the particle (and not, for example, on higher
derivatives of the motion or the local curvature), the subtraction in
Eq.~(\ref{subtract}) will be finite as long as the magnitudes of the
four-accelerations of the two particles are equal and we identify the
local spacetime neighborhoods in such a way that the four-velocities
and four-accelerations, the geodesic distances from the world lines,
and the parallel transport defined by $g_{aa'}$ all coincide up to
$\order[r^0]$. Given points $P$ and $\Ptil$ on two world lines such
that $a^a a_a = \atil^a \atil_a$, we can achieve this by identifying
the spacetime neighborhoods of $P$ and $\Ptil$ with their respective
tangent spaces $T_{P}$ and $T_{\Ptil}$ via the exponential
map\footnote{The exponential map identifies $v^a \in T_{P}$ with the
spacetime point which lies unit affine parameter along the geodesic
generated by $v^a$.}, and then identifying $T_{P}$ and $T_{\Ptil}$ via
any linear map which takes $u^a$ to $\util^a$ and $a^a$ to
$\atil^a$. Under this identification, it is clear that
four-velocities, four-accelerations, and geodesic distances will
coincide exactly, so we need only check that parallel transport will
also agree up to the appropriate order.

One way to see this is to write out Eq.~(\ref{gradphipm}) explicitly
in coordinates adapted to our identification map, so that each point
in the neighborhood of $P$ is mapped to the point with the same
coordinates in the neighborhood of $\Ptil$. (Using such coordinates,
our map identifies a vector field in the neighborhood of $P$ with the
vector field in the neighborhood of $\Ptil$ having the same coordinate
components.) One such coordinate system is Riemann normal
coordinates\footnote{In order to construct Riemann normal coordinates
for a neighborhood of point $P$, identify points in the neighborhood
with points in $T_P$ via the exponential map, and then pick any
orthonormal basis for $T_P$.}. In these coordinates, the coordinate
components of $g_{a'a}$ are given by
\begin{equation}
\gbar_{\al\be}
  = g_{\al\be}
    + \frac{1}{6} r^2 R_{\al\ga\be\de} \rhat^\ga \rhat^\de
    + \order[r^3].
\label{gbarexp}
\end{equation}
(We have dropped the primed indices completely since expression
relates components rather than tensors.) Comparing this to
Eq.~(\ref{gradphipm}), we see that $\gbar_{\al\be}$ simply acts as the
identity at this order in $r$. [The term $- (1/6) q R_{\al\ga\be\de}
\rhat^\ga \rhat^\de \rhat^\al$, which arises from the multiplication
of the $r^{-2}$ term in Eq.~(\ref{gradphipm}) and the $r^2$ term in
Eq.~(\ref{gbarexp}), vanishes by the symmetries of the Riemann
tensor.]  Therefore, the divergent terms will indeed cancel under the
identification we have described. This provides the basis of our
crucial Comparison Axiom in the next subsection.

\subsection{The axiomatic approach}
\label{axioms}

We are now prepared to give our prescription for $\fS^a$, the total
scalar force acting on the particle. We have seen that the subtraction
of field gradients in Eq.~(\ref{subtract}) will be finite as long as
the two particles' four-accelerations have the same magnitude and we
identify the spacetime neighborhoods via the exponential map as
described above. We now elevate this property to the status of an
axiom that any prescription for $\fS^a$ must satisfy.

\noindent
{\bf Axiom 1 (Comparison Axiom)} Consider two points,~$P$
and~$\widetilde{P}$, each lying on timelike world lines in possibly
different spacetimes which contain Klein-Gordon fields $\phi$ and
$\phitil$ sourced by particles of charge $q$ on the world lines. If
the four-accelerations of the world lines at~$P$ and~$\Ptil$ have the
same magnitude, and if we identify the neighborhoods of~$P$
and~$\Ptil$ via the exponential map such that the
four-velocities and four-accelerations are identified, then the
difference between the scalar forces~$\fS^a$ and~$\fStil^a$ is given by the
limit as~$r \rightarrow 0$ of the field gradients, averaged over a
sphere at geodesic distance~$r$ from the world line at $P$.
\begin{equation}
\fS^a - \tilde\fS^a = \lim_{r \rightarrow 0}
	q \langle \nabla^a\phi - \nablatil^a\phitil \rangle_r
\end{equation}

Since the Comparison Axiom requires only that the four-accelerations
of the particles agree, we now need only fix the dependence of $\fS^a$
on acceleration in some arbitrary spacetime in order to uniquely
determine $\fS^a$. Motivated by the time-reflection symmetry of the
half-advanced, half-retarded solution for a uniformly accelerating
trajectory in flat spacetime, we impose the following axiom, which
should be familiar from electromagnetism.

\noindent
{\bf Axiom 2 (Flat spacetime axiom)}
If~$(M,g_{ab})$ is Minkowski spacetime, the world line is uniformly
accelerating, and~$\phi$ is the half-advanced, half-retarded
solution, $\phi =\frac{1}{2}(\phip + \phim)$, then~$\fS^a=0$ at
every point on the world line.

We will now show that, if there exists a prescription for $\fS^a$
satisfying these two axioms, it must be unique. Consider a point $P$
on the world line of a scalar particle of charge $q$ in some
spacetime, and let the particle have acceleration $a^a$ at point
$P$. Let $\fS^a$ and $\gS^a$ be two prescriptions for the scalar
force, both satisfying the axioms given above. Now consider a
uniformly accelerating particle with the same charge $q$ and the same
acceleration $a^a$ in a flat spacetime $(\reals^4, \eta_{ab})$, and
construct the half-advanced, half-retarded solution $\phitil =
\frac{1}{2}(\phiptil + \phimtil)$ for this particle. By our
second axiom, we know that $\fStil^a=\gStil^a=0$ at every point $\Ptil$
along the world line of this uniformly accelerating
particle. Therefore, identifying the neighborhoods of $P$ and $\Ptil$
as in the Comparison Axiom above, we have
\begin{equation}
\fS^a - \gS^a
  = (\fS^a - \fStil^a) - (\gS^a - \gStil^a)
  = \lim_{r \rightarrow 0} q \langle \nabla^a\phi - \nablatil^a\phitil\rangle_r
  - \lim_{r \rightarrow 0} q \langle \nabla^a\phi - \nablatil^a\phitil\rangle_r
  = 0.
\end{equation}

This argument establishes uniqueness, but it also demonstrates
existence by providing a prescription which is guaranteed to satisfy
the axioms. Namely, given a point $P$ along the world line of a scalar
particle with charge $q$ in any spacetime, we simply construct the
half-advanced, half-retarded solution $\phitil$ for a uniformly
accelerating particle in flat spacetime with the same charge and
acceleration. The scalar force $\fS^a$ is then given by
\begin{equation}
\fS^a 
  = \lim_{r \rightarrow 0} q \langle \nabla^a\phi
                                     - \nablatil^a\phitil\rangle_r.
\end{equation}
This is the prescription for the total scalar force which we set out
to find at the beginning of this section.

Writing $\phi$ as $\phi = \phiin + \phim$, we can use
Eq.~(\ref{gradphipm}) to turn this prescription into an explicit
formula for $\fS^a$. The result is
\begin{eqnarray}
\fS^{a} 
  &=& q \nabla^{a}\phiin + q^2 \left(
        \frac{1}{3} (\adot^{a} - a^2 u^{a})
      + \frac{1}{6} (R^{ab} u_{b} + R_{bc} u^{b} u^{c} u^{a})
      - \frac{1}{12} R u^{a}
      \right)
\nonumber \\ && \mbox{}
      + \lim_{\epsilon \rightarrow 0}
         q^2 \int_{-\infty}^{\tau - \epsilon} \!\!
               \nabla^a \Gm(z(\tau),z(\tau')) \, d\tau'.
\label{fS}
\end{eqnarray}
This expression, which is the main result of the paper, allows us to
calculate $\fS^a$ for any trajectory $z(\tau)$ in any spacetime. As
stated at the beginning of the section, the physical significance of
this expression is that it should correctly describe the order $q$ and
$q^2$ contributions to the force on a nearly spherical extended body
in the point particle limit.

\section{The equations of motion}
\label{eom}

We now wish to consider the special case in which no non-scalar forces
are present, so that the evolution of the world line $z(\tau)$ is
determined by the scalar field. In the next subsection, we derive
equations of motion for $z(\tau)$ in this case. Then, in the following
subsection, we explore one of the consequences of these equations of
motion: that the mass of particle varies with time.

\subsection{Derivation of the equations of motion}
\label{eomderivation}

Consider once again the extended body described in
Sec.~\ref{force}. In the absence of non-scalar fields, conservation of
stress-energy dictates that
\begin{equation}
\nabla_b \Tbody^{ab} = - \nabla_b \TS^{ab}.
\end{equation}
According to the arguments of Quinn and Wald~\cite{quinnwaldforce}, in
the point particle limit, the center of mass world line $z(\tau)$ will
therefore satisfy
\begin{equation}
u^b \nabla_b (m u^a) = \frac{dm}{dt}u^a + m a^a = \fS^a,
\end{equation}
where $\fS^a$ is the limiting force we derived in
Sec.~\ref{force}. Inserting our expression for $\fS^a$ from
Eq.~(\ref{fS}) and separating the components parallel to $u^a$ and
perpendicular to $u^a$, we have
\begin{eqnarray}
a^a &=& \frac{1}{m}(\fS^a + u^a g_{bc} u^b \fS^c) \nonumber \\
  &=& \frac{q}{m} (\nabla^{a}\phiin + u^a u^b \nabla_b\phiin)
      +\frac{q^2}{m}
       \left(\frac{1}{3} (\adot^{a} - a^2 u^{a})
             + \frac{1}{6} (R^{ab} u_{b} + R_{bc} u^{b} u^{c} u^{a})
       \right)
\nonumber \\ && \mbox{}
      + \lim_{\epsilon \rightarrow 0}
         \frac{q^2}{m} \int_{-\infty}^{\tau - \epsilon} \!\!
         (\nabla^a\Gm
          + u^a g_{bc}u^b \nabla^c \Gm) \, d\tau'
\label{accel}
\end{eqnarray}
and
\begin{equation}
\frac{dm}{d\tau} = - \fS^a u_a
  = - q u^a \nabla_a \phiin
    - \frac{1}{12} q^2 R
    - \lim_{\epsilon \rightarrow 0}
         q^2 \int_{-\infty}^{\tau - \epsilon} \!\!
               u_a \nabla^a \Gm \, d\tau'.
\label{dmdtau}
\end{equation}

We now note three important features of these equations. First, for each
point along the world line, the integrals in these expressions
represent that portion of $\nabla^a \phim$ which arises from source
contributions interior to the past light cone of the point. This
contribution to the force, often called the ``tail term,'' is a direct
consequence of the failure of Huygen's principle in curved spacetime,
and can be understood as the result of scalar radiation backscattering
from the background curvature and re-intersecting the particle world
line. The presence of this tail term is the primary obstacle to
applying these equations in physically realistic situations, since
most methods for calculating the retarded field of an arbitrary world
line irretrievably mix the tail and non-tail portions of the field.

Secondly, we can provide further insight into the nature of
Eq.~(\ref{dmdtau}) by tracing the origin of the Ricci scalar term in
the Hadamard expansion of the field given in Sec.~\ref{force}. This
term arises directly from the $\{V\}_{\tau=\taum} \nabla_a\taum$ term
in Eq.~(\ref{gradphimraw}). In particular, we have
\begin{equation}
\lim_{\tau' \rightarrow \tau} \Gm(z(\tau),z(\tau')) = \frac{1}{12} R,
\end{equation}
so that we can rewrite Eq.~(\ref{dmdtau}) as
\begin{equation}
\frac{dm}{d\tau} = - q u^a \nabla_a (\phiin + \phitail),
\label{dmdtaurewrite}
\end{equation}
where $\phitail$ is defined by
\begin{equation}
\phitail = \lim_{\epsilon \rightarrow 0}
         q \int_{-\infty}^{\tau - \epsilon} \!\! \Gm \, d\tau'.
\end{equation}
The implications of Eq.~(\ref{dmdtaurewrite}) for global energy
conservation are explored by Quinn and
Wald~\cite{quinnwaldconservation}.

Finally, owing to the presence of the Abraham-Lorentz $\adot^a$ term,
these equations share the unphysical ``runaway'' solutions which have
been discussed thoroughly in the electromagnetic case. (See
Jackson~\cite{jackson} for one such discussion.) In order to interpret
these solutions, it is important to remember that we view the force
law given by Eq.~(\ref{fS}) as an approximate expression for the force
on an extended body, valid to $\order[q^2]$, rather than a fundamental
description of a point particle. Therefore, we can eliminate these
unphysical solutions through the reduction of order technique. This
technique is discussed in detail by Flanagan and
Wald~\cite{flanaganwald}, but the basic idea is simple. Recall that we
wish Eq.~(\ref{accel}) to describe the limiting motion of a
one-parameter family of extended bodies in which both the charge and
the mass of the bodies vanish as the parameter goes to zero. For
concreteness, let us assume that the charge and mass are given by
$q=a\epsilon$ and $m=b\epsilon$. In order to apply the reduction of
order technique to Eq.~(\ref{accel}), we simply insert the entire
right side of the equation in place of $a^a$ in the $\adot^a$ and
$a^2 u^a$ terms and discard any resulting terms which are
$\order[\epsilon^2]$ or higher. The result is
\begin{eqnarray}
a^a &=& \frac{q}{m} (\nabla^{a}\phiin + u^a u_b \nabla^b\phiin)
\nonumber \\ && \mbox{}
      + \frac{1}{3}\frac{q^2}{m}
        \left(\frac{q}{m}\left(u^b \nabla_b \nabla^a\phiin
                               + u^a u^b u^c \nabla_b \nabla_c \phiin \right)
              - \frac{q^2}{m^2}
                \left(\nabla^b\phiin \nabla_b\phiin
                      + (u^b \nabla_b \phiin)^2\right) u^{a} \right)
\nonumber \\ && \mbox{}
              + \frac{1}{6} \frac{q^2}{m}
                (R^{ab} u_{b} + R_{bc} u^{b} u^{c} u^{a})
      + \lim_{\epsilon \rightarrow 0}
         \frac{q^2}{m} \int_{-\infty}^{\tau - \epsilon} \!\!
         (\nabla^a\Gm
          + u^a g_{bc}\nabla^b\Gm u^c) \, d\tau',
\label{redaccel}
\end{eqnarray}
which is free of the unphysical runaway solutions.

\subsection{Time variation of the mass}

In stark contrast to the electromagnetic case, $\fS^a$
includes contributions which point along the four-velocity of the
particle, resulting in a time-varying mass. This is not a special
feature of the self force, nor of curved spacetime. Rather, it
reflects a fundamental difference between the two continuum
theories. Consider a small body in Minkowski spacetime with a center
of mass world line $z(\tau)$. The rest mass of such a body is given by
\begin{equation}
m = - \int_\Sigma u_a \Tbody^{ab} \epsilon_{bcde},
\end{equation}
where $u^a$ is the four-velocity of $z(\tau)$ (defined away from the
world line by global parallelism), $\Sigma$ is the surface
perpendicular to $u^a$, and $\epsilon_{abcd}$ is the volume element
compatible with the (flat) metric. Therefore, we have
\begin{eqnarray}
\frac{dm}{d\tau}
  &=& - \frac{d}{d\tau} \int_\Sigma u_a \Tbody^{ab} \epsilon_{bcde} \nonumber\\
  &=& - \int_\Sigma \Lie_w [u_a \Tbody^{ab} \epsilon_{bcde}] \nonumber\\
  &=& - \int_\Sigma u_a \nabla_b \Tbody^{ab} w^c \epsilon_{cdef},
\end{eqnarray}
where $w^a$ is the vector field which connects successive time slices
$\Sigma(\tau)$.  For a body coupled to a scalar field, we have
$\nabla_b \Tbody^{ab} = - \nabla_b \TS^{ab} = \rho \nabla^a \phi$,
so that
\begin{equation}
\frac{dm}{d\tau}
  = - \int_\Sigma \rho u_a \nabla^a \phi w^c \epsilon_{cdef},
\end{equation}
which is clearly, in general, nonvanishing. By contrast, in the
electromagnetic case, we have $\nabla_b \Tbody^{ab} = - \nabla_b
\TEM^{ab} = F^{ab} j_b$, so that
\begin{equation}
\frac{dm}{d\tau}
  = - \int_\Sigma u_a F^{ab} j_b \phi w^c \epsilon_{cdef}.
\end{equation}
For typical models of charged matter, $j^a$ and $u^a$ will
become collinear as we take the point particle limit, and $dm/dt$ will
vanish.

Perhaps because it is tempting to generalize from the more familiar
electromagnetic case, this time variation of the mass in the scalar
case has largely been ignored in the literature. Some authors use the
equation of motion $m a^a = q \nabla^a \phi$ (e.g.,\ Shapiro and
Teukolsky~\cite{shapiroteukolsky}). This equation is clearly
inconsistent, and therefore in general has no solutions, since $a^a$
is perpendicular to the four-velocity while $\nabla^a \phi$, in
general, is not. Others explicitly project $\nabla^a \phi$
perpendicular to the four-velocity as in Eq.~(\ref{accel}) above in
order to obtain the acceleration of the particle, but then simply
ignore the component of $\nabla^a \phi$ which points along $u^a$ and
assume that the mass is constant (e.g., Ori~\cite{ori}). While such an
equation of motion is mathematically consistent, it violates global
conservation of stress-energy. (See Quinn and
Wald~\cite{quinnwaldconservation}.)

In the discussion above, we have motivated our point particle
equations of motion by imposing local stress-energy conservation on
continuum matter and taking the point particle limit, using our axioms
to extract the appropriate finite part of the divergent fields. The
time variation of the mass arises as a direct consequence of this
local stress-energy conservation. In the literature on point
particles, one sometimes sees an alternative derivation which makes no
reference to the continuum theory. Instead, the author defines an
action for the point particle system and then formally minimizes this
action with respect to variations of the fields and the world line in
order to produce equations of motion. For completeness, we give such a
derivation here, paying particular attention to the time dependence of
the particle's mass.

Fix a globally hyperbolic spacetime $(M, g_{ab})$ and two Cauchy
surfaces for the spacetime, $C_1$ and $C_2$. Let $\phi$ be a smooth
scalar field and $z(\tau)$ be a smooth world line in the region $V$
between $C_1$ and $C_2$. We fix the value of $\phi$ and the position
of $z(\tau)$ on $C_1$ and $C_2$ and define the action, $S$, as
\begin{eqnarray}
S &=& \int_V \biggl[\frac{1}{8\pi}
                    \left(g^{ab} \nabla_a \phi \nabla_b \phi\right)
                    + \frac{1}{2} \int m g_{ab} u^a u^b
                                       \delta^4(x-z(\tau)) \, d\tau
                    + \int q \phi \delta^4(x-z(\tau)) \, d\tau
             \biggr] \epsilon_{abcd}.
\end{eqnarray}

Formally minimizing this action with respect to variations of
$\phi$, we arrive at
\begin{equation}
\nabla^a \nabla_a \phi = - 4 \pi \int q\delta^4(x-z(\tau)) \, d\tau,
\label{kgpoint}
\end{equation}
while minimization with respect to variations of $z(\tau)$ yields
\begin{equation}
\frac{dm}{d\tau} u^a + m a^a = q \nabla^a \phi.
\label{eompoint}
\end{equation}
These are the same equations we arrived at by considering the point
particle limit of the continuum theory. Of course, here we have assumed
$\phi$ and $z(\tau)$ to be smooth in order to define the action, while
the solutions of Eq~(\ref{kgpoint}) are clearly
distributional. Therefore, no solutions of these equations
exist. However, we may view this as a formal derivation of our
equations from an action principle.

Note that, if we had assumed from the outset that $m$ was constant,
the only change to the equations would have been to set $dm/d\tau=0$
in Eq.~(\ref{eompoint}). Clearly, the resulting equation is
inconsistent, since $\nabla^a \phi$ does not, in general, point along
the four-velocity. Still, one might wonder, despite the stress-energy
conservation arguments given above, whether the
above action can be modified to produce the equation of motion
\begin{equation}
\frac{dm}{d\tau} u^a + m a^a = q (\nabla^a \phi
                                  + u^a g_{bc} u^b \nabla^c \phi),
\label{projection}
\end{equation}
since this equation would have the immediate consequence that
$dm/d\tau = 0$, as in the electromagnetic
case. Wiseman~\cite{wisemannotes} has considered a large class of
possible coupling terms and has found that, within this class, one
cannot produce Eq.~(\ref{projection}) without introducing a nonlinear
coupling on the right side of Eq.~(\ref{kgpoint}). Based on this work
and the stress-energy considerations discussed above, we conjecture
that there exists no action which produces Eq.~(\ref{projection})
while preserving Eq.~(\ref{kgpoint}).

\section*{Acknowledgements} I would like to thank Robert Wald
for his expert guidance over the course of the last several years. I
would also like to thank Alan Wiseman for many useful
discussions. This research was supported in part by NSF grant
PHY95-14726 to the University of Chicago.

\end{document}